\documentclass[12pt,amsmath,onecolumn,tightenlines,nofootinbib,aps,pra]{revtex4-1}
\usepackage{graphicx}
\usepackage{hyperref}
\usepackage{bm}
\usepackage{etoolbox}
\usepackage{titlesec}
\patchcmd{\section}
{\centering}
{\raggedright}
{}
{}
\patchcmd{\subsection}
{\centering}
{\raggedright}
{}
{}
\titleformat{\section}
{\normalfont\fontsize{15}{15}\bfseries}{\thesection}{1em}{}
\renewcommand\thesection{\arabic{section}}

\newcommand\pubnumber{CIPANP2015-Lamm}
\newcommand\pubdate{\today}

\def\asu{Physics Department\\ Arizona State University, Tempe, Arizona 
85287, USA}
\def\support{\footnote{Work supported by the National Science Foundation under Grant Nos. PHY-1068286 and PHY-1403891.}}

\def\Title#1{\begin{center} {\Large #1 } \end{center}}
\def\Author#1{\begin{center}{ \sc #1} \end{center}}
\def\Address#1{\begin{center}{ \it #1} \end{center}}

\newcommand\pubblock{\rightline{\begin{tabular}{l} \pubnumber\\
         \pubdate  \end{tabular}}}
\newenvironment{Abstract}{\begin{quotation}  }{\end{quotation}}
\newenvironment{Presented}{\begin{quotation} \begin{center} 
             PRESENTED AT\end{center}\bigskip 
      \begin{center}\begin{large}}{\end{large}\end{center} \end{quotation}}

\def\beq{\begin{equation}}
\def\eeq#1{\label{#1}\end{equation}}
\def\eeqn{\end{equation}}

\def\beqa{\begin{eqnarray}}
\def\eeqa#1{\label{#1}\end{eqnarray}}
\def\eeqan{\end{eqnarray}}

\let\bar=\overbar

\def\Dslash{\not{\hbox{\kern-4pt $D$}}}
\def\dslash{\not{\hbox{\kern-2pt $\del$}}}

\def\msb{{\bar{\ssstyle M \kern -1pt S}}}

\begin{document}
\begin{titlepage}
\pubblock

\vfill
\Title{True Muonium: the atom that has it all}
\vfill
\Author{ Henry Lamm\support}
\Address{\asu}
\vfill
\begin{Abstract}
Fundamental physics could be described as having a muon problem. Discrepancies between theory and experiment in a number of muonic measurements ($r_P$, $a_\mu$, $H\rightarrow\mu\tau$, $\mathcal{R}_K$, $\mathcal{R}_{D^*}$) have been observed in the last decade that demand explanation. True muonium, the bound state of a muon and its antiparticle, has the potential to put strong constraints on any beyond standard model physics that affects the muon sector. To study these effects, theoretical predictions and experiment need a, not unreasonable, precision of 100 MHz. Surprisingly, at this level, QCD and electroweak corrections become relevant in true muonium as well. Presented here is a discussion of the novel properties of true muonium from the standard model, potential discovery methods at near future facilities like DIRAC and HPS, and its ability to discriminate between a variety of solutions to the muon problem.
\end{Abstract}
\vfill
\begin{Presented}
Conference on the Intersections of \\Particle and Nuclear Physics\\
Vail, CO, USA,  May 19--24, 2015
\end{Presented}
\vfill
\end{titlepage}
\def\thefootnote{\fnsymbol{footnote}}
\setcounter{footnote}{0}

\section{Introduction}
The discovery of the Higgs boson at the LHC completes the Standard Model.  With it, we have found the only Standard Model source of flavor-dependent interactions.  From this perspective the current state of physics might be described as having a ``muon problem''.  The first, perhaps more curious, part of this problem is why does the Standard Model have muons, and other higher generations, at all.  Along side this, we have the fact that several muon observables are found to have varying levels of disagreement with Standard Model calculations: the anomalous magnetic moment $a_\mu$ at $2.9\sigma$\cite{PhysRevD.73.072003}, the proton's charge radius from muonic hydrogen $r_P(\mu p)$ at $7\sigma$\cite{Antognini:1900ns}, the decay of $D$ and $B$ mesons to leptons at $\approx2\sigma$\cite{Aaij:2014ora,Aaij:2015yra}, and even Higgs decays at $2.5\sigma$\cite{CMS:2014hha}. As has been discussed in many papers, solving the muon problem has strongly discriminates between beyond Standard Model theories\cite{Heeck:2010pg,Jaeckel:2010xx,Batell:2011qq,Batell:2009jf,TuckerSmith:2010ra,Barger:2010aj,Barger:2011mt,Carlson:2012pc,Onofrio:2013fea,Wang:2013fma,Kopp:2014tsa,Gomes:2014kaa,Karshenboim:2010cm,Karshenboim:2010cg,Karshenboim:2010cj,Karshenboim:2010ck,Karshenboim:2011dx,Karshenboim:2014tka,Brax:2014zba,Lamm:2015gka}.

A strong candidate for shedding light on the muon problem is the bound state $(\mu^+\mu^-)$, dubbed ``true muonium''\cite{Hughes:1971}.  Simpler bound states like positronium $(e^+e^-)$, hydrogen, and muonium $(\mu^+e^-)$ have attracted significant attention as testing grounds for precision QED studies\cite{Karshenboim:2005iy}, but are limited in their BSM discovery potential by either the small electron mass or large uncertainties from unknown nuclear structure effects.  In contrast, true muonium has a much larger reduced mass, and its QCD corrections are limited to the better-understood hadronic vacuum polarization effects, due to its leptonic nature.  

Unfortunately, true muonium has yet to be directly observed.  The first reason is the technical difficulty of producing low-energy muon pairs, coupled with the bound state's short lifetime ($\tau\approx$ 1 ps), which presents an interesting challenge to experimenters.  A second, more prosaic, reason true muonium has been neglected is until the $a_\mu$ anomaly, there seemed no reason to expect true muonium would offer any novel physics to justify the experimental effort.

In this talk, we begin by describing the experimental prospects for studying true muonium.  We will then discuss some features of the true muonium system that could be used to explore possible solutions to the muon problem.  

\section{Experiments}
While the lifetime of true muonium is short, the real issue with experimental investigations of it is producing a sufficiently large number of the bound state.  In the past, many  proposed production channels have been discussed:  $\pi p\rightarrow (\mu^+\mu^-)n$\cite{Bilenky:1969zd}, $\gamma Z\rightarrow (\mu^+\mu^-) Z$\cite{Bilenky:1969zd}, $e Z\rightarrow e(\mu^+\mu^-) Z$\cite{Holvik:1986ty,ArteagaRomero:2000yh},  $\mu^+\mu^-\rightarrow (\mu^+\mu^-)$\cite{Hughes:1971}, $e^+e^-\rightarrow (\mu^+\mu^-)$\cite{Moffat:1975uw,Brodsky:2009gx}, $e^+e^-\rightarrow (\mu^+\mu^-)\gamma$\cite{Brodsky:2009gx}, $\eta\rightarrow (\mu^+\mu^-)\gamma$\cite{Nemenov:1972ph,Kozlov:1987ey}, $Z_1Z_2\rightarrow Z_1Z_2(\mu^+\mu^-)$\cite{Ginzburg:1998df}, and $q^+q^-\rightarrow(\mu^+\mu^-)g$ in a quark plasma\cite{Chen:2012ci}.  Some of the more novel methods of utilizing these production channels considered include fixed target experiments\cite{Banburski:2012tk}, Fool's Intersection Storage Rings\cite{Brodsky:2009gx}, and even from astrophysical sources\cite{Ellis:2015eea}.

Currently, the Heavy Photon Search (HPS)\cite{Celentano:2014wya} experiment has plans to search for true muonium \cite{Banburski:2012tk}, and DImeson Relativistic Atom Complex (DIRAC) \cite{Benelli:2012bw} has discussed the possibility of its observation in an upgraded run\cite{dirac}.  Additionally, the DIRAC experiment intends to study the Lamb shift in the $(\pi^+\pi^-)$ bound state using a fixed magnetic field and measuring the decay rate as a function of distance\cite{Nemenov:2001vp}, and the methods developed there could be used to also measure the Lamb shift in true muonium.  In both situations, the true muonium could be traveling at relativistic speeds, and it may be necessary to consider the effect of this boost on the wave functions in order to get precision production and annihilation rates.  Initial calculations have been undertaken in Ref.~\cite{Lamm:2013oga} toward this goal using a toy model.

Another, older method developed for measuring the Lamb shift in hydrogen\cite{Robiscoe:1965zz} exist and has been suggested as a possible way to measure the Lamb shift in true muonium\cite{Brodsky:2009gx}.  In this method, a beam of $2s$ state true muonium would be passed through a magnetic field, resulting in a mixing of the states with the $2p$ state which then decays to the ground state, decreasing the intensity of the beam.  By measuring the beam's intensity as a function of magnetic field, the Lamb shift can be extracted.

In the further future, the development of the Muon Accelerator Program (MAP) at Fermilab opens up the possibility of high intensity beams of $\mu^+$ and $\mu^-$.  These beams are proposed at the source of both high quality neutrino beams and muons for use in a muon collider.  Using these high intensity beams, it could be possibly to create sufficient amounts of true muonium for precision laser spectroscopy and rare decay measurements.  
\section{Spectroscopy}
Predictions for BSM models solving the muon problem generically lead to corrections to the Lamb shift and hyperfine splitting of true muonium as large as $\mathcal{O}(100 $ MHz) (e.g, \cite{TuckerSmith:2010ra}) which is likely an experimentally accessible level.  In addition to relatively large effects, true muonium is a good system for investigating BSM physics because the existence of an annihilation channel implies any new field content effects most observables (e.g. Lamb shift, $1s-2s$ interval, hyperfine splitting, production and decay rates).  This is in contrast to other observables like $a_\mu$ and $(\mu H)$ spectroscopy where individual measurements can be insensitive to different types of particle content (i.e. Pseudoscalar contributions to the Lamb shift in $(\mu H)$ are heavily suppressed)  

$\mathcal{O}(100 $ MHz) corresponds to about $\mathcal{O}(m\alpha^7)$ corrections to the true muonium spectrum; therefore, we must have a standard model prediction of this level.  Currently, the theoretical predictions for the HFS in true muonium are known fully to $\mathcal{O}(m\alpha^5)$ \cite{Jentschura:1997tv,Jentschura:1997ma,Karshenboim:1998am}, with some $\mathcal{O}(m\alpha^6)$ (see \cite{Adkins:2014dva} and references within for a historical review) and $\mathcal{O}(m\alpha^7)$ which are known for positronium that can also be applied\cite{Melnikov:1999uf,Pachucki:1999zz,Kniehl:2000cx,Melnikov:2000zz,Hill:2000zy,Baker:2014sua,Adkins:2014dva,Eides:2014nga,Adkins:2014xoa}. The current theoretical value of the hyperfine splitting in true muonium is\cite{PhysRevD.91.073008}:
\begin{equation}
 \Delta E^{1s}_{\rm hfs}=42330577(800)(1200)\text{ MHz},
\end{equation}
where the uncertainty arises from model-dependent hadronic effects\cite{Jentschura:1997tv} and estimates of missing $\mathcal{O}(m\alpha^6)$ corrections that do not occur in positronium involving virtual electrons and hadrons\cite{PhysRevD.91.073008}.
In order to obtain an $\mathcal{O}(100 $ MHz) calculation of the HFS, we have found in Ref.~\cite{PhysRevD.91.073008} that single $Z$-boson interactions will have to be included.  This means that in addition to the BSM potential of true muonium, very low energy measurements of the weak interaction are possible, including a measurement of the weak charge of the muon.

Other spectroscopic predictions are not currently computed as precisely.  From Ref.~\cite{Jentschura:1997tv}, the predictions for the $1s-2s$ interval and Lamb shift are
\begin{equation}
 \Delta E_{\rm 1s-2s}=2.55(5)\times 10^{11}\text{ MHz},
\end{equation}
\begin{equation}
\Delta E_{\rm Lamb}= 1.3813(14)\times 10^{7}\text{ MHz}.
\end{equation}
It is important to note that while these calculations all require improvement for use in BSM studies, no new theoretical techniques are required; positronium and muonium techniques can be straight-forwardly be applied to improve these predictions.  Most of the unknown corrections arise from virtual electron loops to photon propagators.

\section{Decays}
Since true muonium has an invariant mass about the two-electron threshold, it is possible for triplet states to annihilate to electron pairs ($\tau_{n^3S_1}=1.8n^3$ ps) , while singlet states will predominately decay to two photons ($\tau_{n^1S_0}=0.6n^3$ ps).  In upcoming experiments, these are the most like channels to be measureable.  If high-intensity true muonium experiments are built, it would potentially be possibly to measure more exotic decays, including those of the triplet state to neutrinos.  The leading order decay rates to mono-energetic neutrinos are known:  
\begin{equation}
 \Gamma(1^3S_1\rightarrow \nu_\mu\bar{\nu}_\mu)=\frac{G_F^2\alpha^3m_\mu^5}{24\pi^2}(1+4\sin^2\theta_w)^2\approx10^{-11}\Gamma_{e^+e^-},
\end{equation}
\begin{equation}
\Gamma(1^3S_1\rightarrow \nu_l\bar{\nu}_l)=\frac{G_F^2\alpha^3m_\mu^5}{24\pi^2}(1-4\sin^2\theta_w)^2\approx10^{-14}\Gamma_{e^+e^-}.
\end{equation}
While these rates are still small, they are within the realm of detected rare processes in mesonic decays, unlike positronium, due to the $\propto m_\ell^5$ scaling.  Related to the measurement of neutrino decays is the general subject of invisible decays.  These have been shown in Ref.~\cite{Badertscher:2006fm} to competitively constrain a variety of BSM (e.g. extra dimensions, axions, mirror matter, fractional charges, and other low-mass dark matter models) in positronium.  In true muonium, these rates are also enhanced due to mass scaling and therefore competitive constraints are possible.

\bibliographystyle{apsrev4-1}
\bibliography{wise}
 
\end{document}